\pdfoutput=1
\RequirePackage{ifpdf}
\documentclass{JINST}

\usepackage{graphicx}
\usepackage{subfigure}

\title{Characterization of the \mbox{FE-I4B} pixel readout chip production run for the ATLAS Insertable B-layer upgrade}

\author{Malte Backhaus\thanks{Corresponding author.}~$^a$ on behalf of the \mbox{ATLAS} \mbox{IBL} collaboration\\
\llap{$^a$}Physikalisches Institut der Unversit\"{a}t Bonn,\\
  Nu{\ss}allee 12, 53115 Bonn, Germany\\
  E-mail: \email{backhaus@physik.uni-bonn.de}}

\abstract{The Insertable B-layer (IBL) is a fourth pixel layer that will be added inside the existing ATLAS pixel detector during the long LHC shutdown of 2013 and 2014. The new four layer pixel system will ensure excellent tracking, vertexing and b-tagging performance in the high luminosity pile-up conditions projected for the next LHC run. The peak luminosity is expected to reach $3 \cdot 10^{34}$\,cm$^{-2}$s$^{-1}$with an integrated luminosity over the \mbox{IBL} lifetime of 300\,fb$^{-1}$ corresponding to a design lifetime fluence of $5 \cdot 10^{15}$\,n$_{eq}$cm$^{-2}$ and ionizing dose of 250\,Mrad including safety factors.\\
The production front-end electronics \mbox{FE-I4B} for the \mbox{IBL} has been fabricated at the end of 2011 and has been extensively characterized on diced ICs as well as at the wafer level. The production tests at the wafer level were performed during 2012. Selected results of the diced IC characterization are presented, including measurements of the on-chip  voltage regulators. The \mbox{IBL} powering scheme, which was chosen based on these results, is described. Preliminary wafer to wafer distributions as well as yield calculations are given.}


\begin{document}

\section{The FE-I4B readout electronics}
\label{sec:fe-i4b}
\begin{figure}[htp]
\centering
  \includegraphics[width=0.6\linewidth]{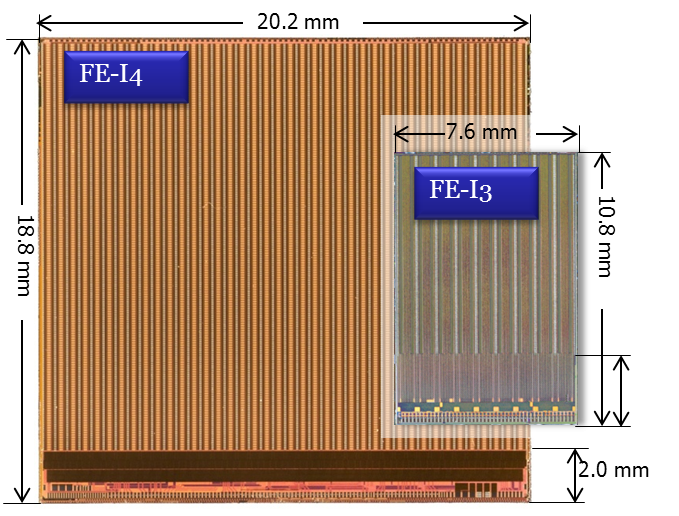}
  \caption{To scale picture of FE-I4 in comparison to FE-I3 readout electronics.}
  \label{fig:fecomp}
\end{figure}

The \mbox{ATLAS} \cite{atlas} Pixel Detector \cite{aad} will be upgraded during the LHC shutdown of 2013/14 (LS1) by the addition of a 4th pixel layer, the so called Insertable B-Layer (\mbox{IBL}) \cite{IBLtdr}. To cope with the limited space available, the \mbox{IBL} will be mounted on a new beam pipe with a smaller radius. The \mbox{IBL} modules will be mounted on 14 staves with a mean radius of only 3.3\,cm with respect to the nominal beam spot. This small radius coupled with the increased high luminosity pile-up results in an increased radiation dose and high hit occupancies. At such low radius, the high hit occupancies would cause increased inefficiencies in the current Pixel Detector readout electronics (FE-I3) \cite{arutinov}. A new readout chip with a different readout architecture was designed for the \mbox{IBL} and future \mbox{ATLAS} Pixel Detector upgrades \cite{barbero} \cite{barbero2} \cite{garcia-sciveres}. The \mbox{IBL} production version of this IC, built in 130\,nm CMOS feature size using thin gate oxide transistors for radiation hardness, is called \mbox{FE-I4B} and is currently the largest chip produced for high energy physics. Its physical size is 20.2 x 18.8\,mm$^2$ with an active area of 20.2 x 16.8\,mm$^2$ and a periphery of 20.2 x 2.0\,mm$^2$, resulting in an active to inactive area fraction of about 90\%, see figure \ref{fig:fecomp}.
The active area holds a pixel matrix organized in 80 columns and 336 rows. Two columns form an architectural unit called a double column (DC). All pixels have a size of 250 x 50\,$\mu$m$^2$ with a two stage amplifier circuit. Four analog pixels share one common digital logic cell. A detailed description of this so called 4-pixel digital region can found in \cite{iblmodules}.
\begin{figure}[htp]
\centering
  \includegraphics[width=0.6\linewidth]{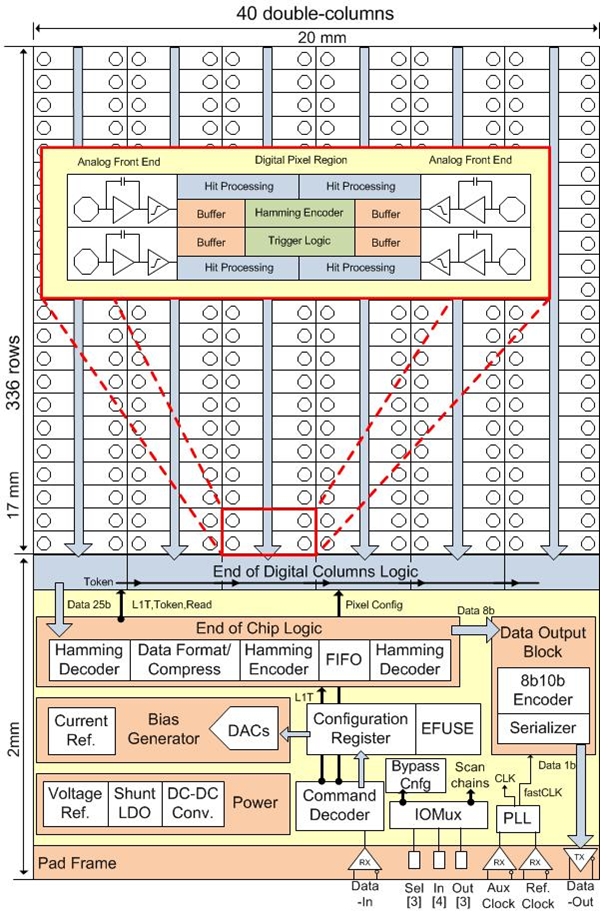}
  \caption{Schematic overview of the FE-I4 readout electronics.}
  \label{fig:fe-i4}
\end{figure}
The hit information is stored in the 4-pixel digital regions of the pixel matrix until arrival of the level 1 trigger. This new architecture avoids copying the information of un-triggered hits to the periphery. This was shown to be the main source of inefficiency in FE-I3 at the expected \mbox{IBL} hit occupancy \cite{arutinov}.\\
FE-I4B is a minor design revision following the successful testing of the previous version, FE-I4A. Most tests carried out on FE-I4A met the requirements for IBL \cite{iblmodules} \cite{barbero3}. While we have checked that the circuits left unchanged from FE-I4A to FE-I4B still perform as expected, this note focuses on those things that were changed. In particular, the implementation of internal power management using built in voltage regulators, and the calibration charge injection circuitry.\\
A simplified block diagram of the chip is shown in figure \ref{fig:fe-i4}. All results presented here on diced ICs as well as at wafer level have been obtained using the USBpix test system \cite{backhaus}.

\section{Main results from FE-I4B IC characterization}
\label{sec:ICresults}
Characterization results of the \mbox{FE-I4B} on diced IC are presented here. Section \ref{subsec:pulser} shows the performance of the test charge injection circuitry and section \ref{subsec:matrix} covers the uniformity of the pixel matrix.

\subsection{Test pulse injection circuitry}
\label{subsec:pulser}
\begin{figure}[htp]
    \centering
    \includegraphics[width=0.6\linewidth]{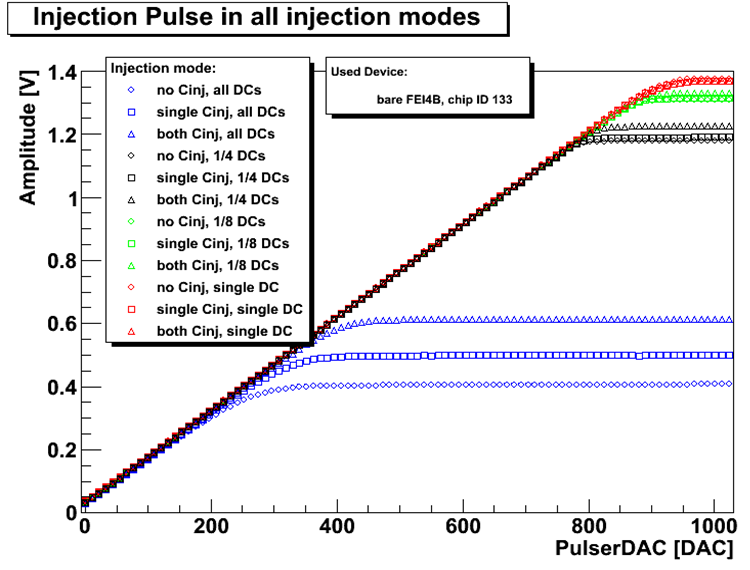}
    \caption{The test charge injection circuitry output pulse amplitude in all injection modes as a function of the corresponding setting for FE-I4B.}
    \label{fig:Pulser}
\end{figure}

A circuit to inject calibration charge into arbitrary pixels is implemented in the FE-I4 design. It works by distributing a voltage step to selectable injection capacitors present in each pixel. This injection circuit had poor performance in FE-I4A,  showing saturation of the achievable voltage step, ranging from a reduced maximum step when a single column was enabled to complete loss of function with all columns enabled. Figure \ref{fig:Pulser} demonstrates the good performance of this block in \mbox{FE-I4B}. The only significant saturation observable in \mbox{FE-I4B} happens when injecting into all 26880 pixels at the same time. It is also possible to inject in every eighth, every fourth double column and into single double columns. All these other modes allow a voltage step across the injection capacitors well above 1.1\,V with good linearity. The saturation was found to be caused by leakage current in the switches used to select the injection capacitors in each pixel. The amount of leakage current depends on which switches are on or off, and figure \ref{fig:Pulser} shows three sets of data for each case of enabled columns because there are three different switch configurations. The performance was improved in FE-I4B by using a low power devices for these switches and by increasing the voltage step drive capability.

\subsection{Pixel matrix}
\label{subsec:matrix}
\begin{figure}[htp]
\centering
    \subfigure[]{
        \includegraphics[width=0.47\linewidth]{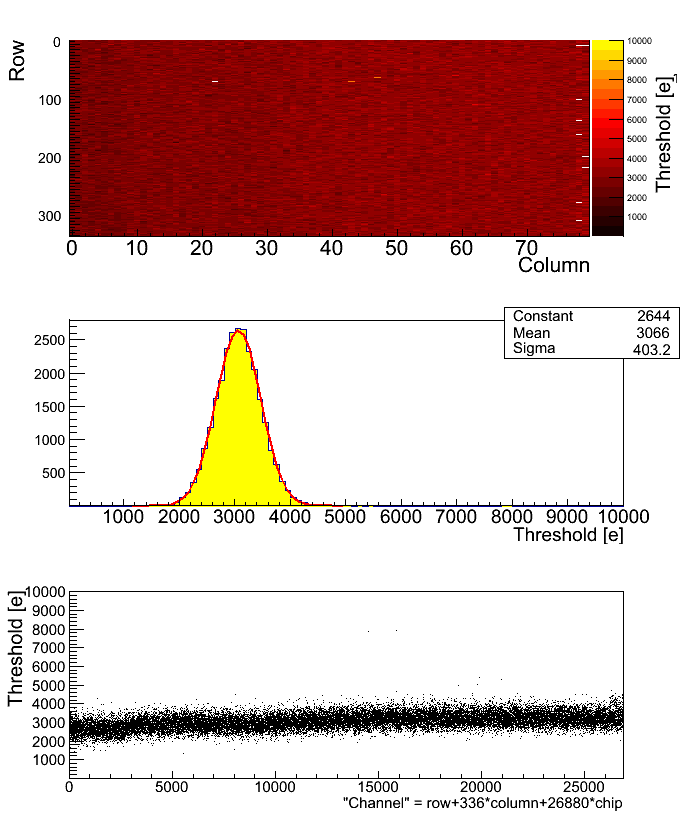}
        \label{fig:MatrixB}
    }
    \subfigure[]{
        \includegraphics[width=0.47\linewidth]{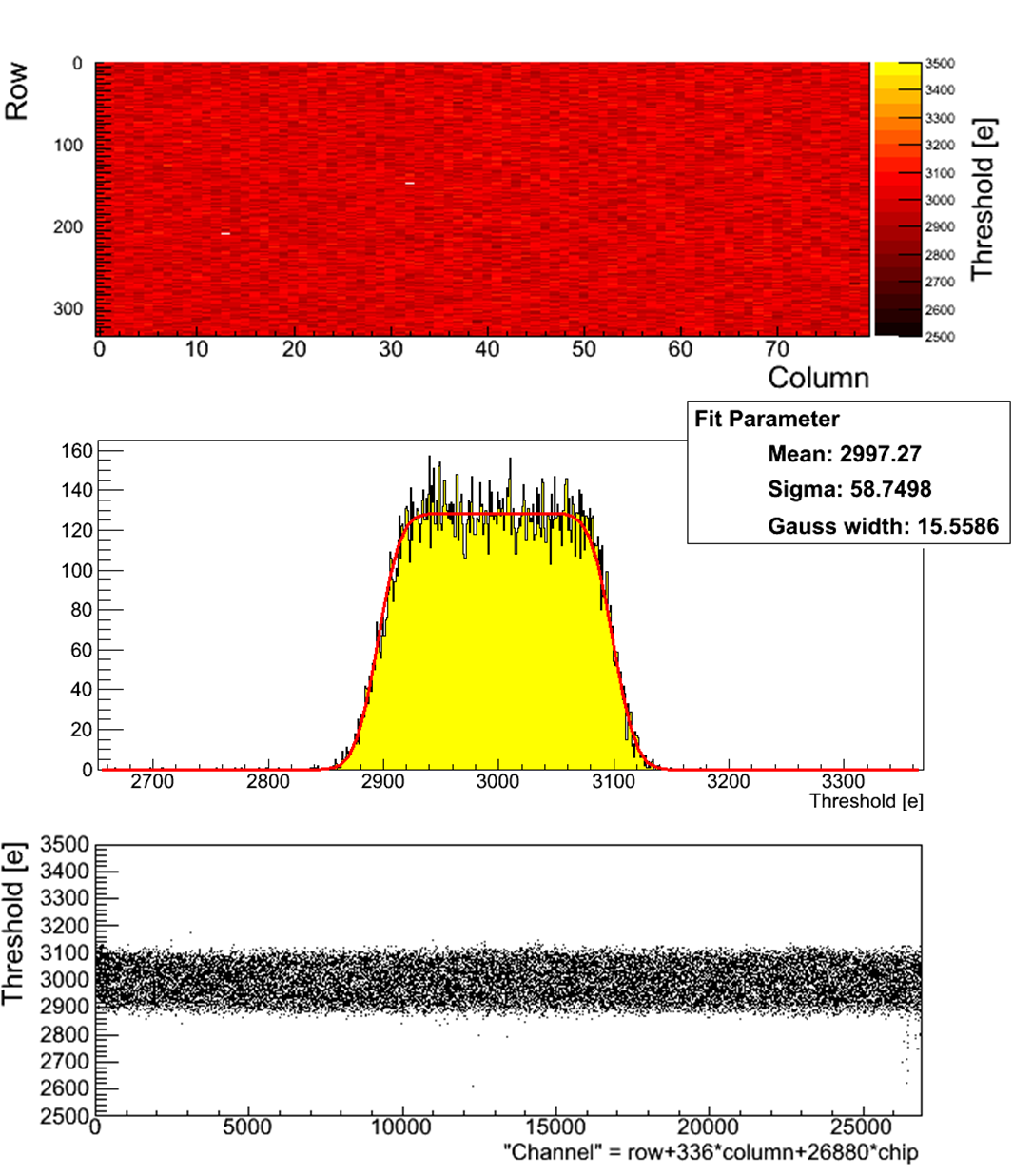}
        \label{fig:ThresholdTuned}
    }
    \caption{An un-tuned threshold map recorded with \mbox{FE-I4B} showing the good uniformity of the pixel matrix (a) and a tuned threshold distribution (b).}
\end{figure}

In the FE-I4A chip different variants of the pixel design were used in some columns in order to measure performance options. These variants included different types of configuration memory cells that must be single event upset (SEU) tolerant, different types of feedback capacitors, and different discriminator designs. As the \mbox{FE-I4B} is the production chip for the \mbox{IBL}, all pixels are now identical, using the variant with the best SEU performance, metal-metal feedback capacitor, and a classic discriminator design. The un-tuned threshold map is sensitive to any non-uniformities in the pixel matrix. The un-tuned threshold map of a \mbox{FE-I4B} is shown in figure \ref{fig:MatrixB}. This map demonstrates the good threshold uniformity of the FE-I4B pixel matrix. On this specific IC only five out of 26880 pixels show a threshold outside of five standard deviations and for ten pixels the measurement of the threshold failed.\\
A threshold tuning DAC (TDAC) is implemented at the pixel level to allow pixel by pixel threshold adjustment. The result of a tuning procedure performed on a \mbox{FE-I4B} is displayed in figure \ref{fig:ThresholdTuned} demonstrating the successful adjustment of the threshold to the target value of 3000 electrons which was chosen here. The expected threshold distribution after tuning is uniform and centered around the target value and with the width of the TDAC step (LSB). As the TDAC step width varies slightly from pixel to pixel, the edges of the uniform distribution are convoluted with a gaussian in the fit function. The standard deviation of this threshold distribution is about 60 electrons after tuning, compared to the standard deviation of the un-tuned distribution of about 400 electrons. Indistinguishable tuning results are obtained in the entire temperature range -40$^{\circ}$C to +40$^{\circ}$C and after proton irradiation up to 300\,Mrad.

\section{The IBL powering scheme}
\label{sec:powering}
\begin{figure}[htp]
\centering
  \includegraphics[width=1.0\linewidth]{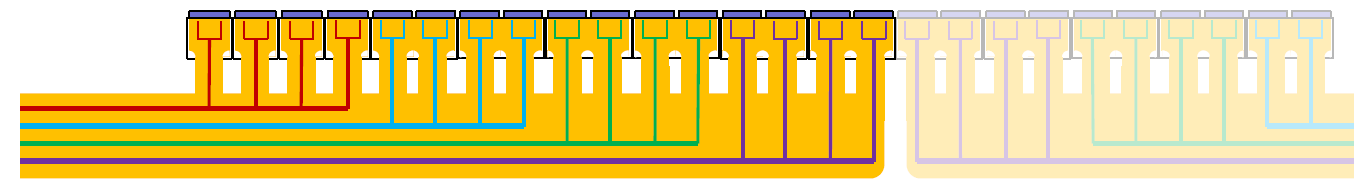}
  \caption{Overview of the module position and low voltage connection of an IBL half stave.}
  \label{fig:stave}
\end{figure}

The \mbox{IBL} staves are split electrically into two half staves. Each half stave holds six double chip modules using a single planar silicon sensor \cite{wittig} provided by CiS\footnote{CiS Forschungsinstitut für Mikrosensorik und Photovoltaik GmbH, Konrad-Zuse-Straße 14, 99099 Erfurt, Germany.} connected to two \mbox{FE-I4B} readout ICs. The double chip modules are mounted towards the middle of the stave. Additionally, four single chip modules using 3D silicon sensors \cite{pellegrini} \cite{dalla} from one of the two providers CNM\footnote{Centro Nacional de Microelectronica (CNM-IMB-CSIC), Campus Universidad Autonoma de Barcelona, 08193 Bellaterra (Barcelona), Spain.} or FBK\footnote{Fondazione Bruno Kessler (FBK), Via Sommarive 18, 38123 Povo di Trento, Italy.} connected to one \mbox{FE-I4B} chip, are mounted at the outside of each half stave. Electrically, four \mbox{FE-I4B} chips are connected to one \mbox{IBL} power supply channel regardless of module type. Each of the Front-End chips holds two on-chip so-called Shunt-LDO (low drop out) regulators \cite{karagounis} \cite{gonella} to generate the analog and digital internal voltages. An overview of the half stave and the \mbox{IBL} power groups is shown in figure \ref{fig:stave}.
The on-chip LDOs are operated in partial shunt mode. This means they are operated as usual LDO as long as the current consumption is above an adjustable minimum input current. If below, an additional current is shunted by the regulators to ground. This operation mode does not increase the power consumption of the Front-End chip as long as its current consumption in working conditions is above the shunt current. The advantage of this mode is the reduction of the transients in comparison to the pure LDO mode in case of load current fluctuations. This will happen in case of configuration of the Front-End chips or accidental configuration loss. The reference voltages needed for the operation of the two LDOs are generated on-chip.

\subsection{The reference voltage circuitries}
\label{subsec:vrefs}
\begin{figure}[htp]
\centering
  \includegraphics[width=0.8\linewidth]{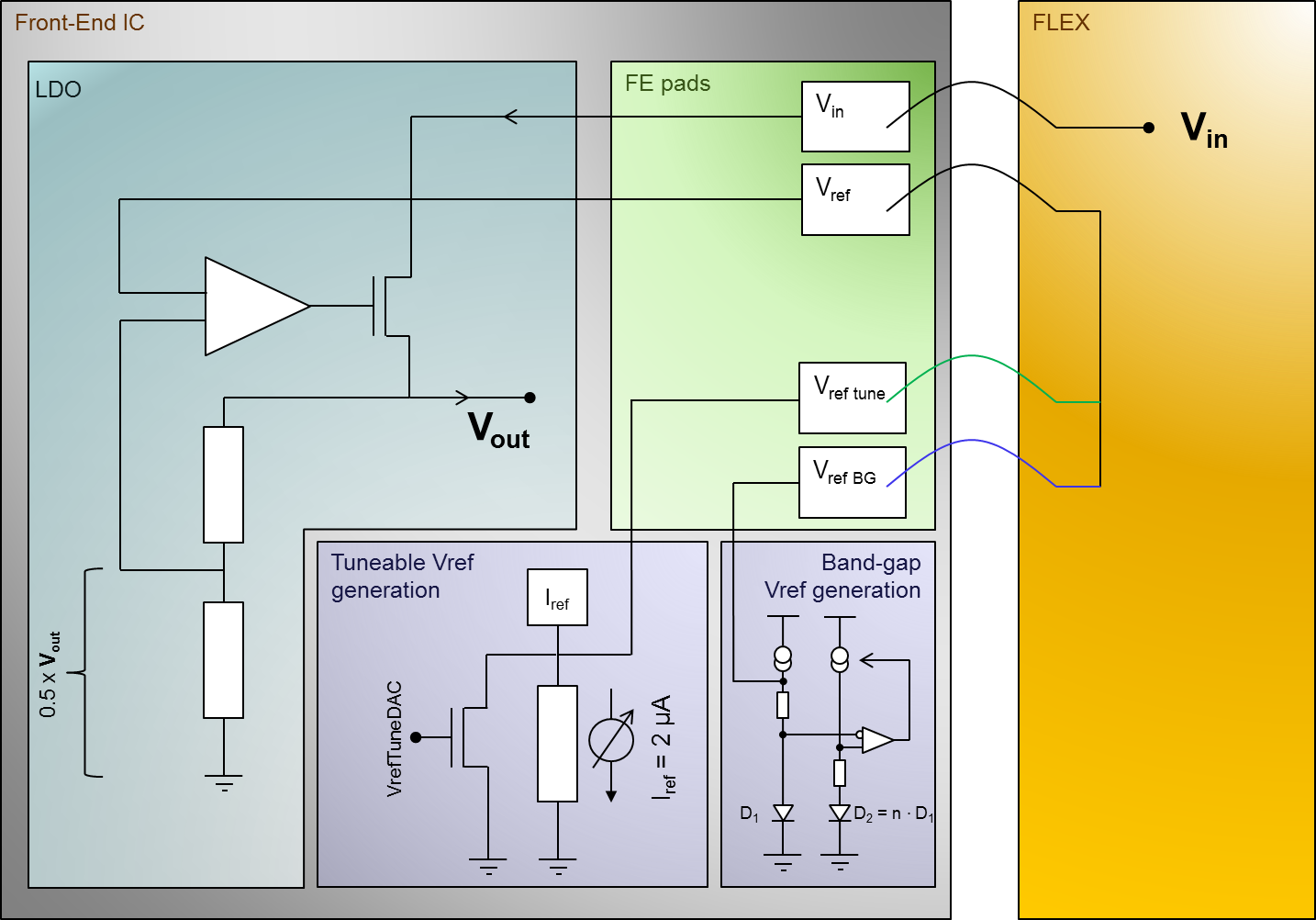}
  \caption{Simplified schematics of the on-chip low-dropout regulators and reference voltage connection options.}
  \label{fig:ldos}
\end{figure}

The FE-I4 chip contains a master current reference, nominally 2\,$\mu$A, from which all internal biases and DACs are fed \cite{gromov_current}. This reference was present in FE-I4A and carried over to FE-I4B. However, this reference is designed for 1.5\,V rail operation and therefore must be powered from the output of a voltage regulator. This presents a startup challenge in the case one wishes to use this reference to control the output of the built-in voltage regulators. A start-up circuit has been introduced in FE-I4B to make this possible. Additionally, new band-gap voltage reference circuits \cite{gromov_band-gap} have been added in FE-I4B for use as optional voltage regulator references. These circuits are designed for 2.5V rail operation and therefore can be powered from the same unregulated voltage feeding the internal regulators. As explained with the results presented below, the final choice for IBL operation is to use the parallel combination of new band-gap reference and existing current reference for the analog regulator reference, which powers the current reference rail, and the current reference alone for the digital regulator.\\\\
The left most box in figure \ref{fig:ldos} shows simplified schematics of the on-chip LDOs. The LDO compares the reference voltage sourced from a wire bond pad to 0.5 times the output voltage and adjusts the output voltage accordingly. Therefore the potential connected to the reference voltage pad should be 0.5 times the required output voltage. There are two independent on-chip reference voltage generation circuitries implemented for each of the two LDOs. One is the above mentioned band-gap reference with fixed output voltage, the other converts the global reference current of 2\,$\mu$A into a voltage using a resistor in parallel to a variable current sink. The sinked current can be adjusted using an on-chip DAC and therefore this voltage reference is called tunable voltage reference. For flexibility one can wire bond the input to the LDOs reference voltage to the band-gap based reference voltage output or to the tunable reference voltage. Both reference voltage blocks have been characterized.

\subsubsection{Characteristics of the band-gap reference voltage}
\label{subsubsec:Vbg}
\begin{figure}[htp]
\centering
  \includegraphics[width=0.45\linewidth]{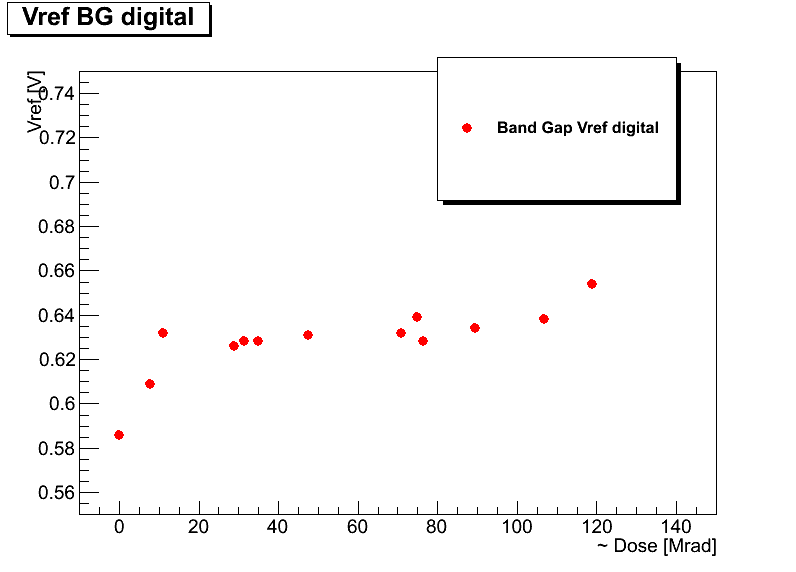}
  \caption{The output voltage of the band gap based reference voltage measured during the 2011 irradiation campaign at Los Alamos.}
  \label{fig:VbgDose}
\end{figure}

The performance of the voltage reference after irradiation is crucial. Therefore \mbox{FE-I4B} bare ICs have been irradiated in Los Alamos irradiation facility using 800\,MeV protons. The output voltage of the band gap based voltage reference has been measured during the irradiation and is plotted as a function of the total ionizing dose in figure \ref{fig:VbgDose}. It has been observed that the band-gap based reference voltage increases with dose. This could endanger the IC after irradiation if the analog input voltage of the IC increases above 1.6\,V. Consequently this reference voltage can not be used by itself for the analog LDO.

\subsubsection{Characteristics of the tunable reference voltage}
\label{subsubsec:Vtune}
\begin{figure}[htp]
\centering
    \subfigure[]{
        \includegraphics[width=0.45\linewidth]{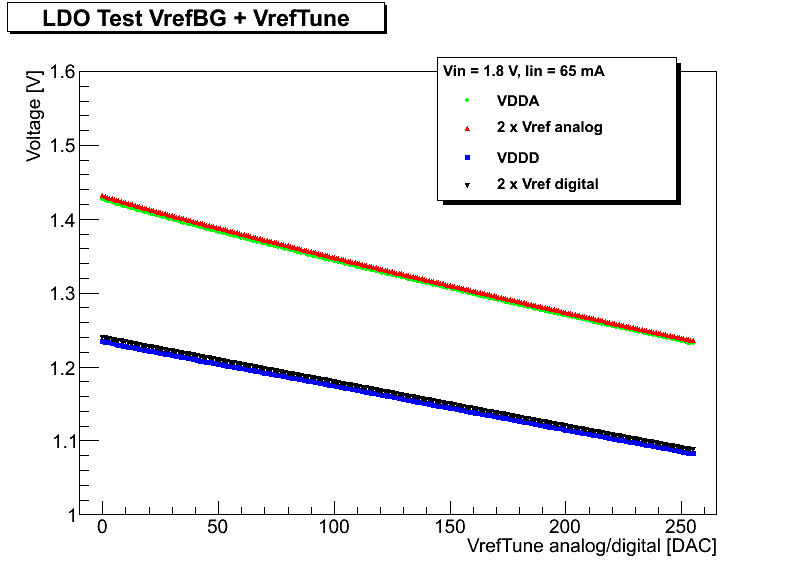}
        \label{fig:VtuneRange}
    }
    \subfigure[]{
        \includegraphics[width=0.45\linewidth]{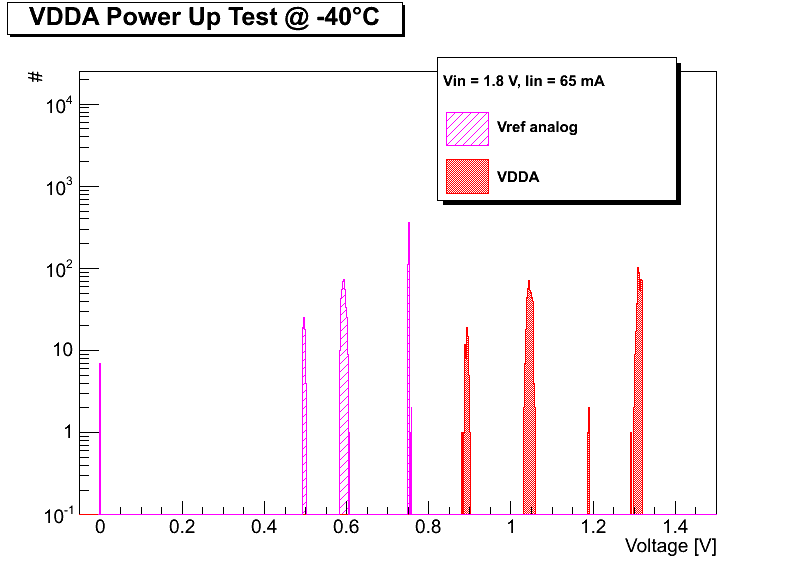}
        \label{fig:VtunePowerUp}
    }
    \caption{Characteristics of the tunable reference voltage. The regulator output voltage as a function of the tune DAC (a) and the result of startup reliability tests at -40 degrees Celsius (b).}
\end{figure}

The tunable reference voltage has a good dynamic range and linearity, see figure \ref{fig:VtuneRange}. However, recall that the start-up is now a concern and a start-up circuit has been added. If the start-up is not fully reliable under all conditions, then also the adjustable reference could not be used by itself to control the analog regulator. The start-up behavior is shown in figure \ref{fig:VtunePowerUp}, where the reference voltage for the analog LDO and the resulting analog LDO output voltage were measured for 1000 power-cycles at -40$^{\circ}$C. A large  fraction of start-up cycles results in too low reference voltage and therefore also LDO output voltage. The behavior becomes reliable for all chips tested above -10$^{\circ}$C, but the minimum reliable temperature varies from chip to chip, and in any case the operation must be qualified down to -40$^{\circ}$C.

\subsection{The IBL reference voltage connection scheme}
\label{subsec:IBLVrefConnection}
\begin{figure}[htp]
\centering
    \subfigure[]{
        \includegraphics[width=0.45\linewidth]{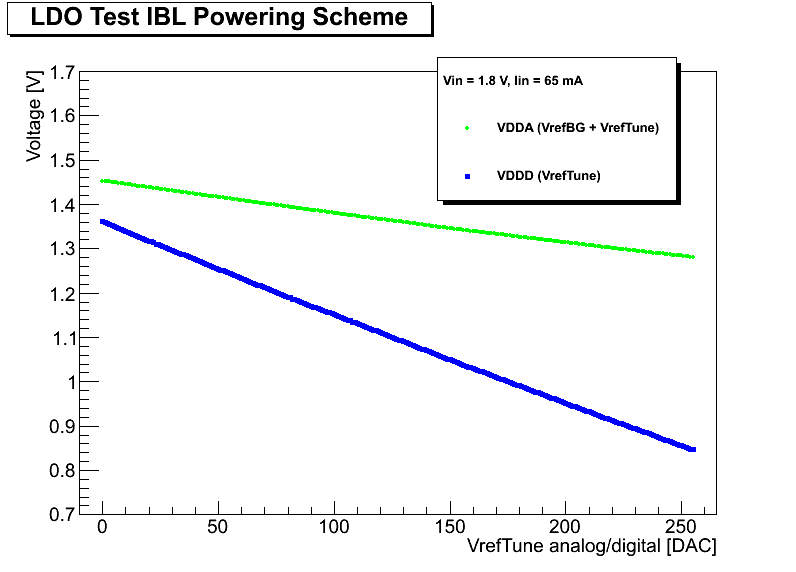}
        \label{fig:VtuneRangeIBL}
    }
    \subfigure[]{
        \includegraphics[width=0.45\linewidth]{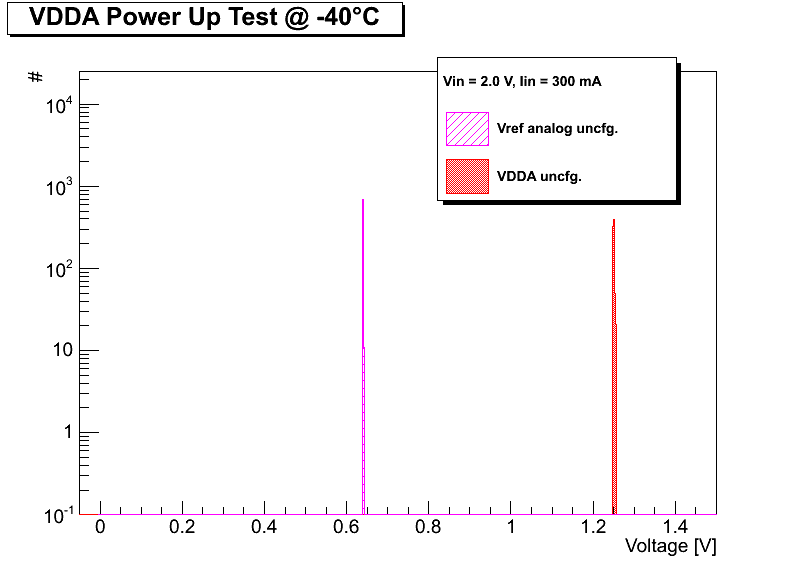}
        \label{fig:VpowerUpIBL}
    }
    \caption{The regulator output voltage as a function of the tune DAC using the \mbox{IBL} powering scheme (a) and the result of startup reliability tests at -40 degrees Celsius of the analog supply voltage in \mbox{IBL} powering scheme (b).}
\end{figure}

A reference voltage connection scheme that provides reliable power-up in a large temperature range while keeping the tunability of the LDO output voltage for both LDOs (and therefore safe operation after heavy irradiation) has been achieved. The regulator outputs of the analog and digital regulators in this powering scheme are shown in figure \ref{fig:VtuneRangeIBL} and the power-up reliability of this connection scheme is demonstrated in figure \ref{fig:VpowerUpIBL}. For the analog regulator the tunable reference voltage and the band-gap reference output are tight together. This provides the benefit of higher startup current for the reference current and results in safe power-up of the analog regulator. The increased value of the band-gap reference voltage with dose can be compensated due to the kept tunability. The design of both references is compatible with parallel connection. For the digital regulator it is sufficient to use the tunable reference voltage only, because the reference current generation block is powered from the analog regulator. This increases the dynamic range of the digital regulator output voltage.

\section{Wafer Level production tests}
\label{sec:wafer}
\begin{figure}[htp]
\centering
  \includegraphics[width=0.45\linewidth]{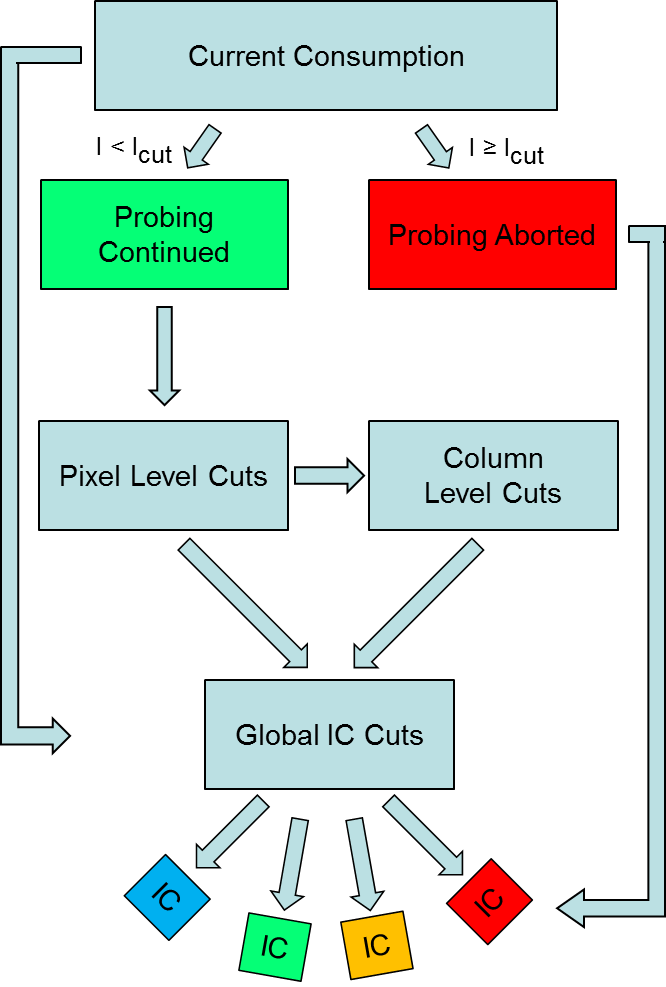}
  \caption{The cut flow of the wafer probing analysis program. The boxes represent different cut levels and arrows illustrate communication directions.}
  \label{fig:cutFlow}
\end{figure}

The wafer level production QA program concentrates on functionality tests of the ICs and measures all IC characteristics that are not accessible after module assembly, such as the charge calibration linearity and offset. Additionally, digital tests such as scan chain tests are performed for all big digital blocks in the ICs periphery. A sophisticated data analysis and cut program is used for automated IC classification. The cut flow of this program is shown in figure \ref{fig:cutFlow}.
The only condition resulting in an abortion of the test run is a significantly high current consumption of the IC after startup. A digital supply current above 350\,mA or an analog supply current exceeding 300\,mA aborts the probing of the IC to protect the probe needles.\\
All pixel level tests such as digital and analog performance tests influence the pixel level cuts. The analog and digital functionality of each pixel is extensively tested. The response to digital and analog injection as well as cross-talk between neighbor pixels are measured. Threshold and noise scans measure the analog performance. The end result is a total number of pixels failing any cut. Up to 53 pixels (0.2\,\% of the chip) are allowed to fail in any cut for an IC suitable for IBL production. The pixel level cuts can translate to column level cuts if the number of failing pixels exceeds the allowed number of broken pixels within one column, which is used in classifying test grade ICs. In addition to pixel and column cuts,  all global IC characteristics such as power consumption, charge calibration or mean noise level feed into the global IC cuts.\\
Finally four different states are assigned to the ICs: green chips are ICs which can be used for the \mbox{IBL} production, yellow ICs are not optimal ICs which can still be used for further R\&D such as future sensor concept characterization, red ICs are non-recoverable and a special state ''blue'' is assigned to any IC for which the software was unable to apply reasonable cuts and therefore human interaction and data crosscheck is needed. It is possible to assign any state from ''blue'' after careful data crosscheck.\\
Several examples of IC measurements and resulting chip to chip distributions are presented in the following sections.

\subsection{The injection capacitance measurement at wafer level}
\label{subsec:WaferCapCalib}
\begin{figure}[htp]
\centering
    \subfigure[]{
        \includegraphics[width=0.45\linewidth]{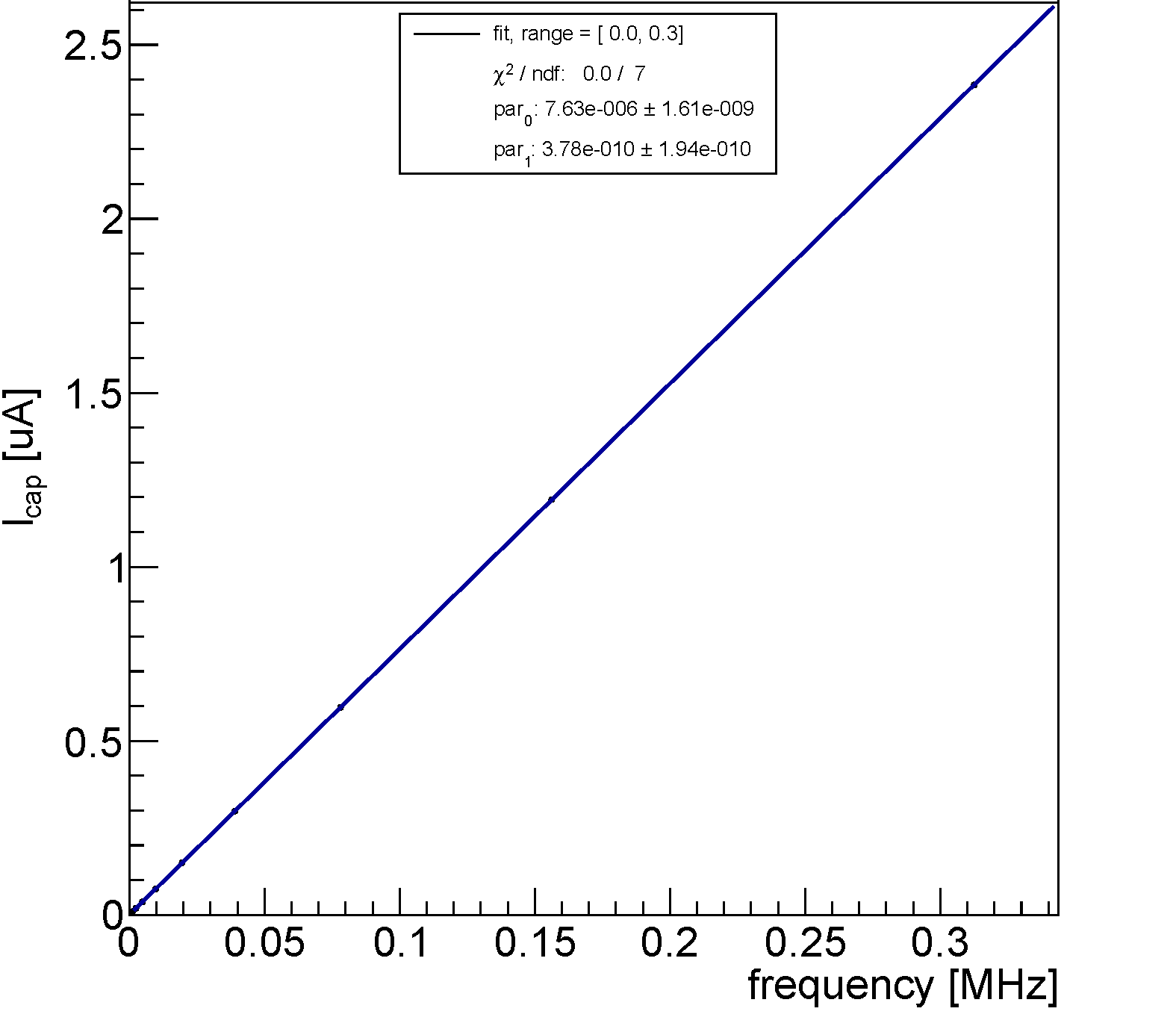}
        \label{fig:Icap}
    }
    \subfigure[]{
        \includegraphics[width=0.45\linewidth]{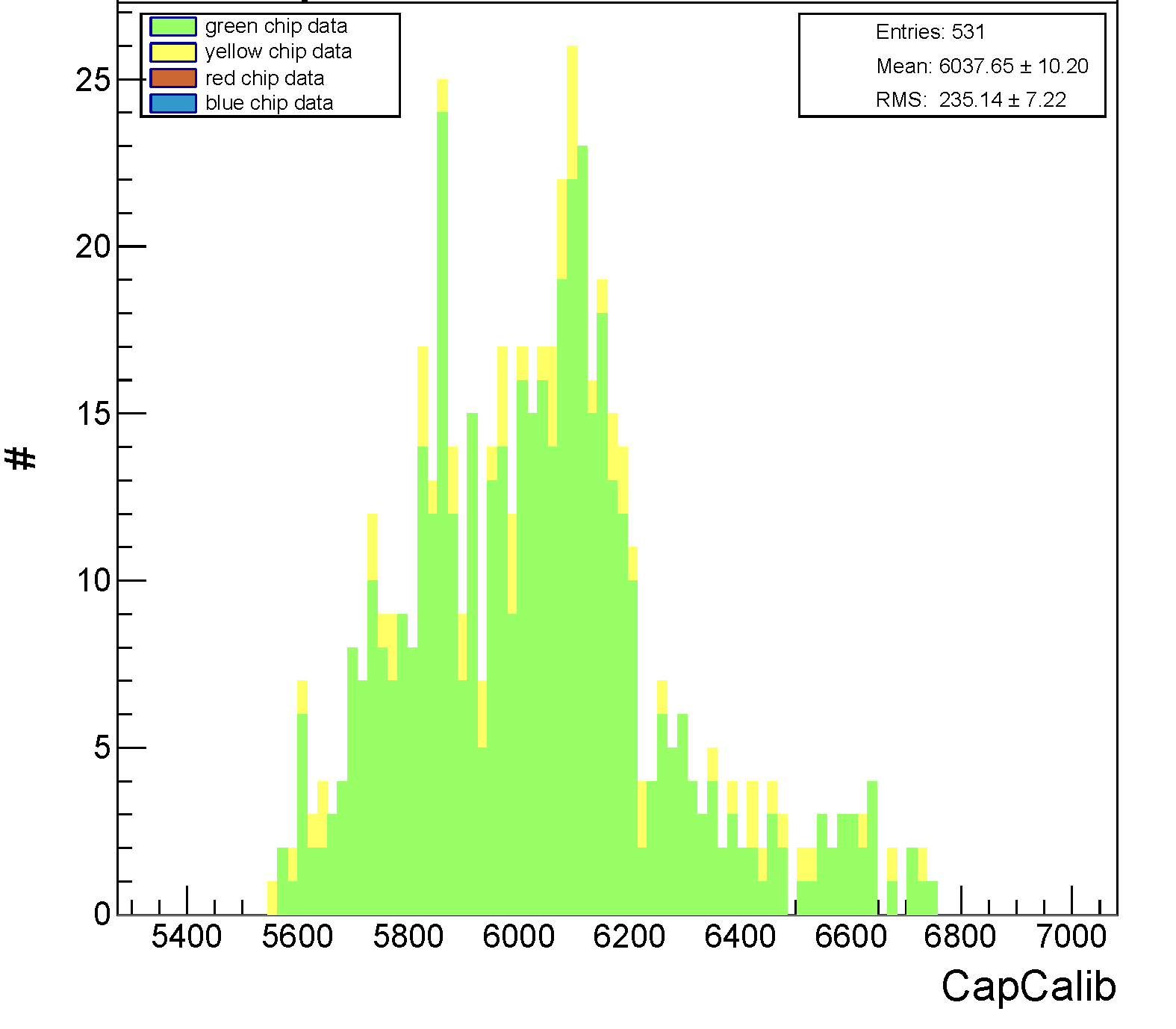}
        \label{fig:CapCalibDist}
    }
    \caption{The single IC measurement of the injection capacitance (a) and the chip to chip distribution for ten wafers in aF (b).}
\end{figure}

A dedicated circuit to measure the average value of representative injection capacitors has been included in FE-I4B. This circuit is not accessible once modules have been assembled. Therefore the test charge injection capacitance must be measured during the wafer level test to achieve a proper charge calibration for each IC. A voltage is applied across an array of 1000 replica injection capacitors in parallel and is switched with variable frequency between input and ground using non-overlapping clocks. The average current is measured, resulting in the linear behavior shown in figure \ref{fig:Icap}. The test charge injection capacitance is computed from the slope of the linear regression. The resulting capacitance distribution for all ICs of ten wafers is shown in figure \ref{fig:CapCalibDist}. The mean test charge injection capacitance is 6.0\,fF with a RMS of  0.235\,fF, which is in good agreement with the simulation results of 5.7\,fF. The observable level chip to chip spread and simulation uncertainty are expected. All injection capacitances are measured to be between 5.6\,fF and 6.8\,fF.

\subsection{Threshold and noise distribution at wafer level}
\label{subsec:WaferThreshNoise}
\begin{figure}[htp]
\centering
    \subfigure[]{
        \includegraphics[width=0.45\linewidth]{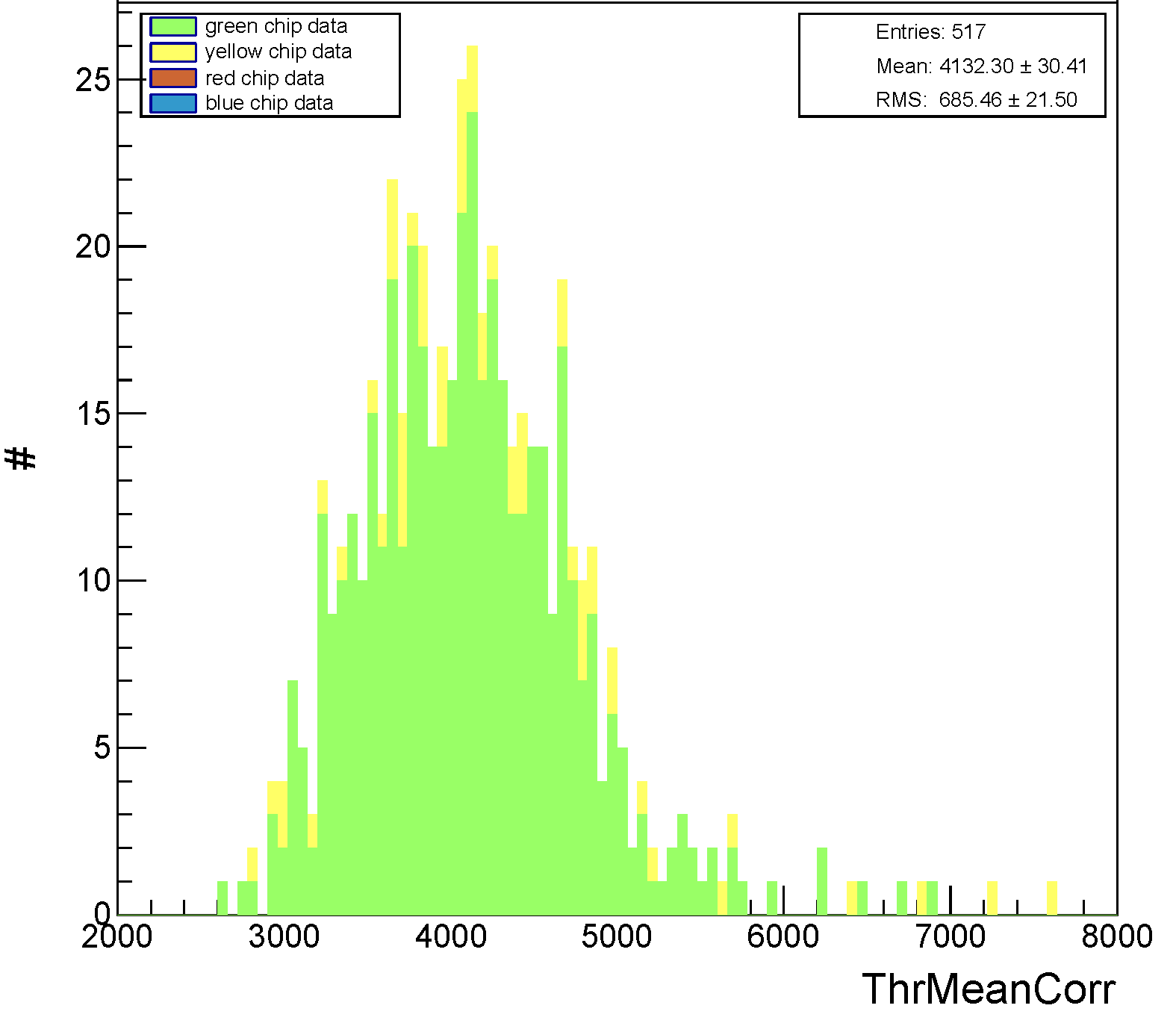}
        \label{fig:ThresholdDist}
    }
    \subfigure[]{
        \includegraphics[width=0.45\linewidth]{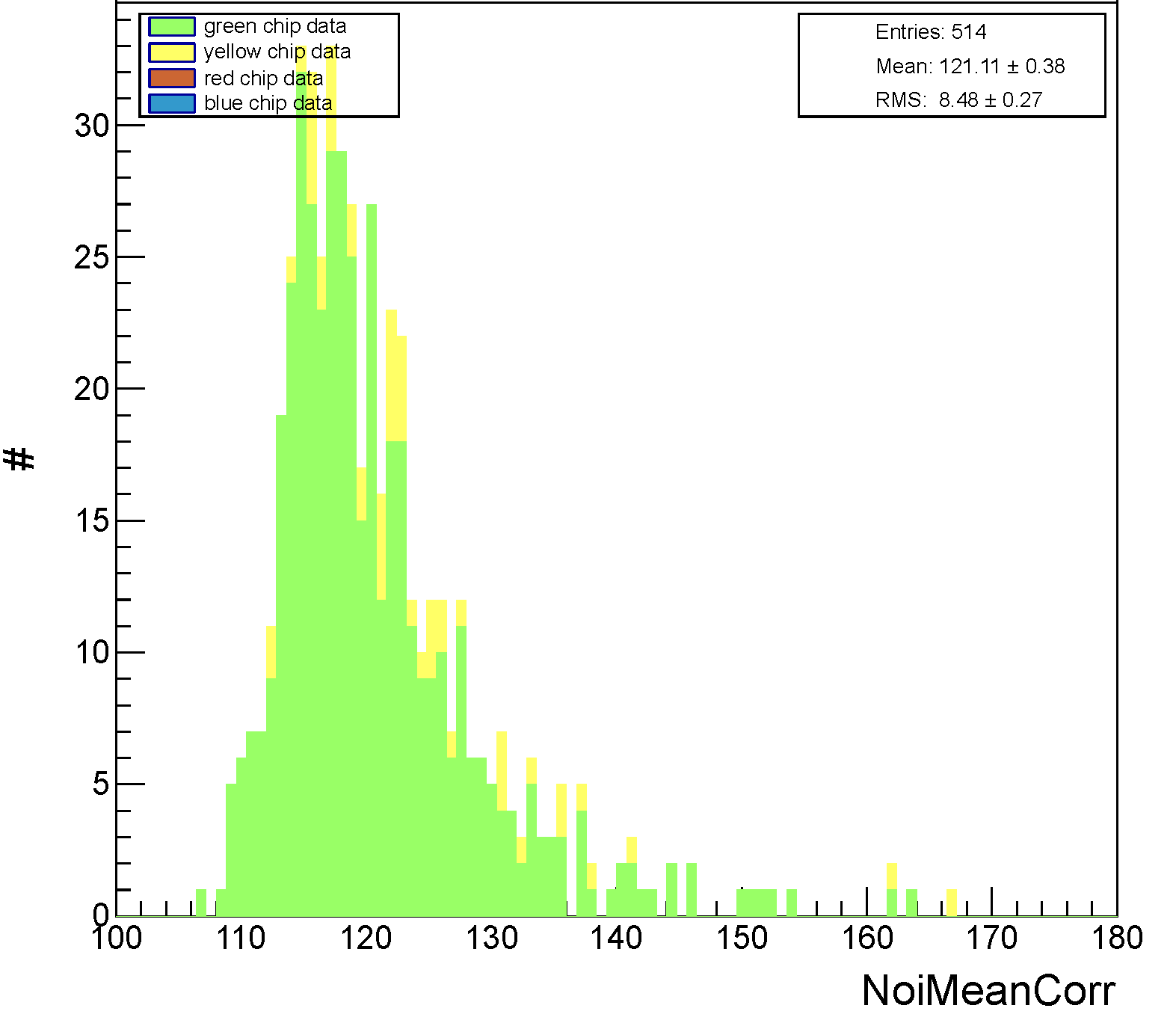}
        \label{fig:NoiseDist}
    }
    \caption{The threshold (a) and noise (b) distribution in electrons for ten wafers.}
\end{figure}

A threshold and noise measurement is performed on each IC. The thresholds are not tuned during the wafer level tests and therefore a wide chip to chip distribution as seen in figure \ref{fig:ThresholdDist} is expected. The noise is not expected to be considerably influenced by the mean threshold of the IC and therefore the noise distribution is expected to be centered around the mean noise value of ~125 electrons as measured on bare \mbox{FE-I4A} prototype ICs \cite{backhaus2}. Indeed the chip to chip distribution of the mean noise value measured on ten wafers peaks at 121 electrons with a RMS of 8.5 electrons.

\subsection{Wafer level test summary}
\label{subsec:WaferSummary}
\begin{figure}[htp]
\centering
  \includegraphics[width=0.8\linewidth]{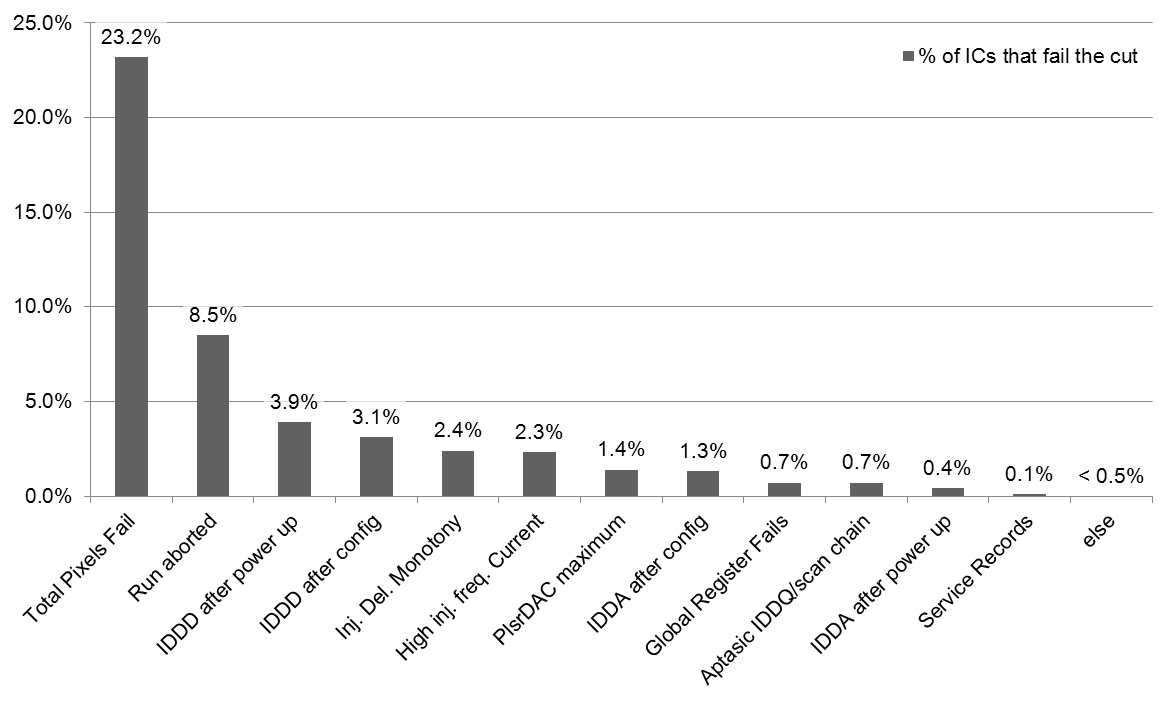}
  \caption{Fraction of ICs failing per acceptance cut.}
  \label{fig:yieldKillers}
\end{figure}

\begin{figure}[htp]
\centering
  \includegraphics[width=0.45\linewidth]{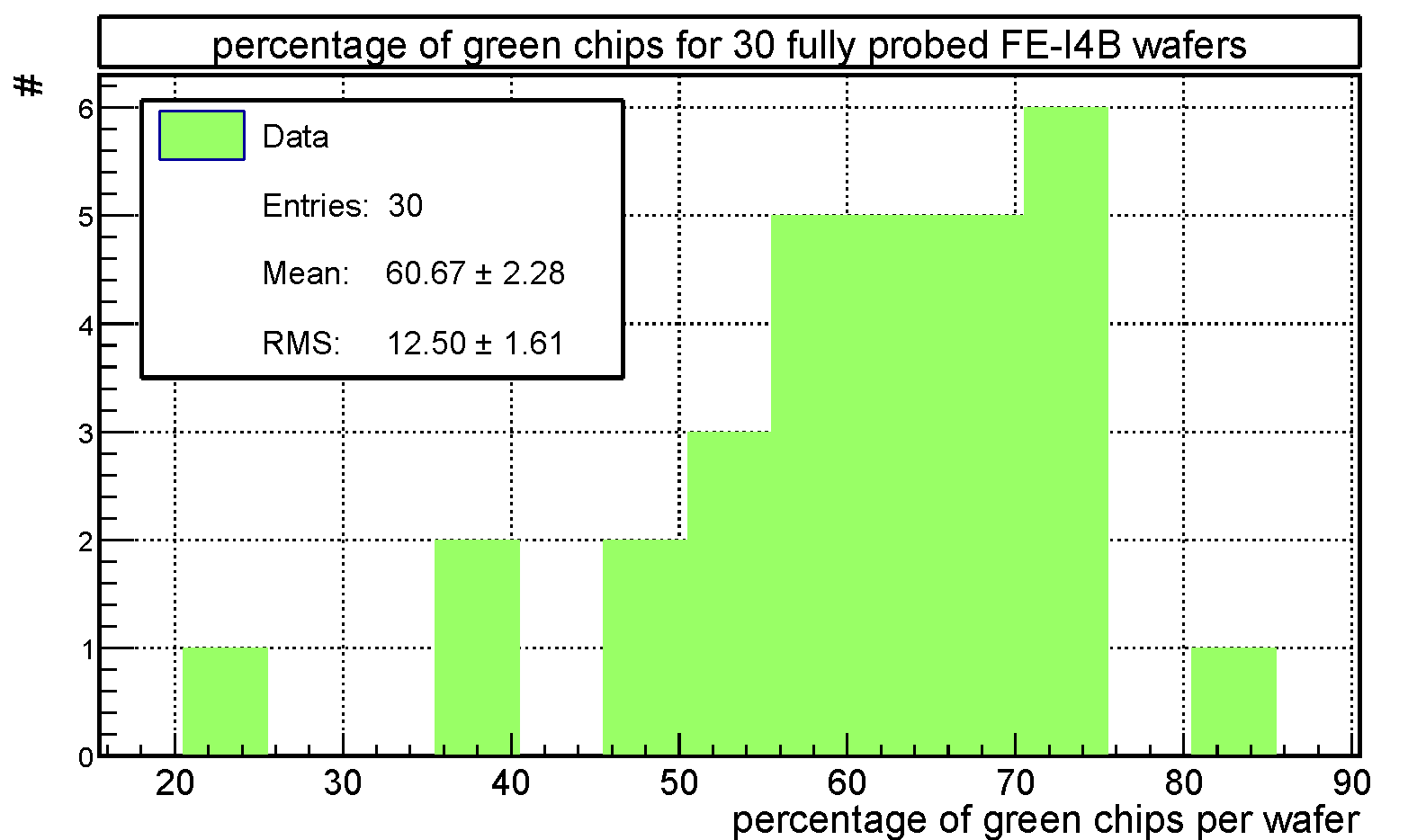}
  \caption{The yield per wafer distribution for 30 fully probed \mbox{IBL} wafers.}
  \label{fig:yield}
\end{figure}

The fraction of ICs failing as a function of the tested feature is shown in figure \ref{fig:yieldKillers}. The biggest influence on the yield is the cut on the total number of failing pixels. This is a very strict cut allowing only 0.2\% of the pixels to show any error in any test. The number of allowed pixels failing was inherited from the current \mbox{ATLAS} Pixel Detector production.\\
The second largest amount of ICs excluded from \mbox{IBL} production comes from high power consumption at various test stages. As a high current consumption is the only condition triggering an abortion of the test run the bin called ''Run aborted'' with 8.5\% of failing ICs contains the ICs with very high current consumption. Together with the third and fourth largest bins and ''IDDA after config'' and ''IDDA after power up'', which contain ICs with completed probing runs but significantly high current consumption, the cuts on the power consumption exclude another 17.2\% of ICs from \mbox{IBL} production.\\
The preliminary mean yield for 30 tested out of 90 produced wafers is 60 $\pm$ 2\%. The percentage of green ICs per wafer for these 30 wafers is shown in Fig.\ref{fig:yield}, showing a coherent region of wafers with 48\% of green ICs to 73\% of green ICs. Three wafers show significantly low yield and one wafer shows a high yield of 83\% green ICs.

\section{Conclusions}
\label{sec:conclusions}
The \mbox{IBL} production readout electronics, \mbox{FE-I4B}, has been characterized at bare IC as well as at wafer level production tests. The test pulse injection circuitry shows good performance. Maximal voltage steps above 1.1\,V are possible in all modes except injecting to all double columns at the same time. The pixel matrix shows good uniformity before and after tuning.\\
A powering option of the on-chip low dropout regulators has been developed that provides reliable power-up performance over the full temperature range while keeping the tunability of both the analog and digital internal voltages of the \mbox{FE-I4B}. This option will be used for \mbox{IBL} operation.\\
The chip to chip distribution of the injection capacitance measured during the wafer level production tests shows a mean value of 6.0\,fF which is in good agreement with the simulations. The mean noise of 121\,electrons measured at wafer level is also in the expected range. On 30 tested wafers 60\,\% of the IC's do fulfill the strict cut criteria for IBL usage.



\end{document}